\documentclass[12pt, prd, showpacs]{revtex4}
\usepackage{amssymb}
\usepackage{amsmath}

\input{tcilatex}

\begin{document}

\title{Confined Penrose process and black-hole bomb}
\author{O. B. Zaslavskii}
\affiliation{Department of Physics and Technology, Kharkov V.N. Karazin National
University, 4 Svoboda Square, Kharkov 61022, Ukraine}
\email{zaslav@ukr.net }

\begin{abstract}
We consider the decay of \ a particle with some energy $E_{0}>0$ inside the
ergosphere of a black hole. After the first decay one of particles with the
energy $E_{1}<0$ falls towards a black hole while the second one with $%
E_{2}>E_{0}\,\ $moves in the outward direction. It bounces back from a
reflecting shell and, afterwards, the process repeats. For radial motion of
charged particles in the Reissner-Nordst\"{o}m metric, the result depends
strongly on a concrete scenario. In particular, an indefinitely large growth
of energy inside a shell is possible that gives rise to a black-hole bomb.
We also consider a similar multiple process with neutral particles in the
background of a rotating axially symmetric stationary black hole. We
demonstrate that, if particle decay occurs in the turning point, a
black-hole bomb in this case is impossible at all. For a generic point
inside the ergoregion, there is a condition for a black-hole bomb to exist.
It relates the ratio of masses before and after decay and the velocity of a
fragment in the center of mass frame.
\end{abstract}

\maketitle

\section{Introduction}

There exist two universal mechanisms of extraction of energy from black
holes. The first one is the Penrose process. In the original form, it was
found \cite{pen} for rotating black hole backgrounds. If in a space-time
there exists a region where negative Killing energies $E<0$ are possible,
then a parent particle 0 can decay to two fragments \ 1 and 2 in such a way
that $E_{1}<0$ while $E_{2}>E_{0}$, so amplification of the original energy
occurs. The ergosphere is realized in the region when the component of the
metric tensor $g_{tt}$ changes the sign as compared to infinity ($t$ is time$%
)$. Later on, it turned out that a similar process occurs in the background
of the Reissner-Nordstr\"{o}m (RN) metric as well \cite{ruf}. In contrast to
the aforementioned rotating case, now the ergosphere is not a pure geometric
entity but depends on the parameters of a particle. For further development
of the Penrose process with charged particles see, e.g. refs. \cite{den} - 
\cite{luca}.

From the other hand, there exists a wave analogue of the Penrose process.
This is a so-called superradiance \cite{st1}, \cite{st2} (for a recent
review and list of references see \cite{pani}). The Penrose process is
realized with particles, superrdiance - with waves. In turn, superradiance
leads to the possibility of one more interesting physical effect -
black-hole bomb. It occurs if a black hole is surrounded by a reflecting
shell and a wave bounces back in such a way that the process repeats
endlessly, the energy is accumulated without bound giving rise to an
explosion. This was shown for reflection by a concave mirror in \cite{press}
and for a convex one in \cite{vb}. (The relation between the existence of
such a bomb and instabilities due to the properties of quasinormal modes was
discussed in \cite{jb}.)

Strange as it may seem, the question about possibility of a black-hole bomb
on the basis of processes with particles was posed only quite recently \cite%
{conf}. The authors considered motion of charged particles in the background
of the Reissner-Nordstr\"{o}m metric with decay of a parent particle to two
fragments with consequent reflection from a shell backward. The event of
decay was chosen to occur in the turning point. The authors demonstrated
convincingly that a black-hole bomb for such a scenario is impossible. In
this sense, the difference between the standard Penrose process and its
multiple version in a confined system is blurred.

However, such a situation is not universal. In the present paper we show
that for another scenarios within the same system a black-hole bomb is
indeed possible. Apart from this, we consider a similar problem for rotating
axially symmetric metrics. It turns out that if the process occurs
(similarly to the RN case) in the turning point, the energy remains bounded,
no black-hole bomb is possible. However, for more general scenarios, this
may happen and we indicate a simple condition on the parameters of a system,
necessary for an indefinitely large growth of energy.

We use the system of units in which fundamental constants $G=c=1$.

\section{Particle motion in the Reissner-Nordstr\"{o}m metric: basic equators%
}

Let us consider the RN metric. It has the form%
\begin{equation}
ds^{2}=-dt^{2}f+\frac{dr^{2}}{f}r^{2}d\omega ^{2}\text{,}
\end{equation}%
where%
\begin{equation}
f=1-\frac{2M}{r}+\frac{Q^{2}}{r^{2}}\text{,}
\end{equation}%
$M$ is the black hole mass, $Q$ being its electric charge. We take $Q>0$.

For simplicity, we consider pure radial motion, so equations of motion read%
\begin{equation}
m\dot{t}=\frac{X}{f}\text{,}
\end{equation}%
\begin{equation}
p^{r}\equiv m\dot{r}=\sigma P\text{, }P=\sqrt{X^{2}-m^{2}f},  \label{P}
\end{equation}%
\begin{equation}
X=E-q\varphi \text{,}  \label{x}
\end{equation}%
where dot denotes differentiation with respect to the proper time. Here, $E$
is the particle energy, $m$ being its mass. The electric Coulomb potential%
\begin{equation}
\varphi =\frac{Q}{r}\text{.}
\end{equation}%
The forward-in-time condition implies $\dot{t}>0$, whence%
\begin{equation}
X>0  \label{X}
\end{equation}%
outside the horizon. Hereafter, we will use notations%
\begin{equation}
\varepsilon =\frac{E}{m},\text{ }\tilde{q}=\frac{q}{m}\text{.}
\end{equation}

The system can have a turning point $r_{t}$, where $P=0.$ Then,%
\begin{equation}
r_{t}=\frac{1}{\varepsilon ^{2}-1}(\varepsilon \tilde{q}Q-M\pm \sqrt{C})%
\text{,}  \label{tp}
\end{equation}%
\begin{equation}
C=(M-\varepsilon \tilde{q}Q)^{2}+(1-\tilde{q}^{2})Q^{2}(\varepsilon ^{2}-1).
\end{equation}%
In what follows we will be interested in the case when a particle has the
energy $E>m$, so $\varepsilon >1$.

\section{General scenario of decay}

Let a parent particle 0 decay in the point with $r=r_{0}$ to particles 1 and
2. We assume the conservation laws%
\begin{equation}
E_{0}=E_{1}+E_{2}\text{,}  \label{e12}
\end{equation}%
\begin{equation}
q_{0}=q_{1}+q_{2}\text{,}
\end{equation}%
\begin{equation}
p_{0}^{r}=p_{1}^{r}+p_{2}^{r}\text{,}
\end{equation}%
whence%
\begin{equation}
X_{0}=X_{1}+X_{2}\text{.}  \label{x12}
\end{equation}

Then, using these equations and (\ref{P}), (\ref{x}) one can obtain

\begin{equation}
X_{1}=\frac{1}{2m_{0}^{2}}\left( X_{0}b_{1}+P_{0}\delta \sqrt{d}\right) ,
\label{e1}
\end{equation}%
\begin{equation}
X_{2}=\frac{1}{2m_{0}^{2}}\left( X_{0}b_{2}-P_{0}\delta \sqrt{d}\right) ,
\label{e2}
\end{equation}%
\begin{equation}
b_{1,2}=m_{0}^{2}\pm (m_{1}^{2}-m_{2}^{2}),  \label{del}
\end{equation}%
where $i=0,1,2$, 
\begin{equation}
d=b_{1}^{2}-4m_{0}^{2}m_{1}^{2}=b_{2}^{2}-4m_{0}^{2}m_{2}^{2},  \label{D}
\end{equation}%
$\delta =\pm 1$. For radial momenta one obtains

\begin{equation}
P_{1}=\left\vert \frac{P_{0}b_{1}+\delta X_{0}\sqrt{d}}{2m_{0}^{2}}%
\right\vert \text{,}  \label{p1}
\end{equation}%
\begin{equation}
P_{2}=\left\vert \frac{P_{0}b_{2}-\delta X_{0}\sqrt{d}}{2m_{0}^{2}}%
\right\vert .  \label{p2}
\end{equation}

Alternatively, one can take advantage of the results already obtained in
eqs.~(19) --- (30) of~\cite{centr}. We use particle labels 1 and 2 instead
of 3 and 4 respectively in~\cite{centr}.

The direction of motion is characterized by a quantity $\sigma $, where $%
\sigma =+1$ for motion in the outward direction and $\sigma =-1$ for the
inward case. We are mainly interested in the situation, when particle~0
moves towards a black hole, from large radii to smaller ones. Then, at least
one of particles falls in a black hole.

\section{Particular scenario of decay\label{rn}}

We will mainly concentrate on the case when decay occurs in the turning
point for all three particles, so $P_{i}=0$ for $i=0,1,2$ and $r_{0}=r_{t}$,
so that 
\begin{equation}
X_{i}=m_{i}\sqrt{f(r_{0})}\text{,}  \label{xt}
\end{equation}%
$i=0,1,2$. It is clear from (\ref{x12}) that this requires 
\begin{equation}
m_{0}=m_{1}+m_{2}\text{.}  \label{m12}
\end{equation}%
Then, it follows from (\ref{P}), (\ref{del}), (\ref{D}) that $%
b_{1}=2m_{1}m_{0}$, $b_{2}=2m_{0}m_{2}$, $d=0$.

Let, in addition, a reflecting shell be placed at some point $r_{B}>r_{0}$.
Particle 0 decays to 1 and 2, particle 1 moves towards a black hole, while
particle 2 moves in the outward direction, bounces back from the shell and
decays again to particles 3 and 4, etc. Similarly to \cite{conf}, we
consider a scenario in which decay happens in the same point $r_{0}$. We can
use the formulas from the previous sections in which substitutions $%
0\rightarrow 2n$, $1\rightarrow 2n+1$ and $2\rightarrow 2n+2$ are made, $%
n=0,1,2...$ Then, using (\ref{x}) and (\ref{xt}) we have

\begin{equation}
E_{2n+1}=m_{2n+1}\sqrt{f(r_{0})}+\frac{q_{2n+1}Q}{r_{0}}\text{,}  \label{e+1}
\end{equation}%
\begin{equation}
E_{2n+2}=m_{2n+2}\sqrt{f(r_{0})}+\frac{q_{2n+2}Q}{r_{0}}\text{.}  \label{e+2}
\end{equation}

The energy $E_{2n+1}<0$, provided $q_{2n+1}=-\left\vert q_{2n+1}\right\vert $%
, where $\left\vert q_{2n+1}\right\vert >q_{2n+1}^{\ast }$,%
\begin{equation}
q_{2n+1}^{\ast }=\gamma m_{2n+1}\text{, }\gamma =\frac{r_{0}}{Q}\sqrt{%
f(r_{0})}\text{.}
\end{equation}

As $m_{2n}<m_{0}$ is finite, the final result depends crucially on $q_{2n}$.
If $\lim_{n\rightarrow \infty }q_{n}=q_{\infty }<\infty $, $E_{2n}$ is
finite as well. This is just what happens in the scenario considered in \cite%
{conf} (we call it Scenario 1). Below we compare two qualitatively different
types of scenario.

\subsection{Scenario 1}

To make presentation self-contained, in this Section we outline briefly the
scenario studied in \cite{conf}. One can assume that splitting of the charge
to two fragments occurs in such a way that

\begin{equation}
q_{2n+1}=-(1+\Delta )q_{2n+1}^{\ast }\text{, }q_{2n+2}=q_{2n}-q_{2n+1}\text{,%
}
\end{equation}%
where $\Delta $ is some constant. We assume that in each act of decay

\begin{equation}
m_{2n+1}=\alpha _{1}m_{2n}\text{, }m_{2n+2}=\alpha _{2}m_{2n},  \label{m12n}
\end{equation}%
where%
\begin{equation}
\alpha _{1}+\alpha _{2}=1\text{.}  \label{a}
\end{equation}%
As a result,%
\begin{equation}
m_{2n}=\alpha _{2}^{n}m_{0},  \label{m2n}
\end{equation}%
\begin{equation}
m_{2n+1}=\alpha _{1}\alpha _{2}^{n}m_{0}\text{.}  \label{m2n1}
\end{equation}%
To avoid confusion, we use here another notations as compared to \cite{conf}%
, quantities without tilde and with it are interchanged. Then, repeating
transformations carried out in \cite{conf}, we arrive at the expressions%
\begin{equation}
q_{2n+1}=-(1+\Delta )\frac{m_{2n+1}\sqrt{f}}{Q}r_{0}=-m_{0}(1+\Delta )\gamma
\alpha _{1}\alpha _{2}^{n}
\end{equation}%
\begin{equation}
E_{2n+1}=-m_{2n+1}\sqrt{f}\Delta \text{,}
\end{equation}%
\begin{equation}
\varepsilon _{2n+1}=-\sqrt{f(r_{0})}\Delta .
\end{equation}%
This means that $\frac{q_{2n+1}}{m_{2n+1}}=const$ does not depend on $n$ as
well as $\varepsilon _{2n+1}$. Then, it follows from the conservation laws
that%
\begin{equation}
q_{2n}=q_{0}+(1+\Delta )\gamma (1-\alpha _{2}^{n})\text{,}  \label{q2n}
\end{equation}%
\begin{equation}
E_{2n}=E_{0}+m_{0}\gamma \Delta (1-\alpha _{2}^{n})\frac{Q}{r_{0}}.
\label{e2n}
\end{equation}%
Eq. (\ref{q2n}) corresponds to eq. (3.32) of \cite{conf} and eq. (\ref{e2n})
corresponds (in our notations) to eq. (3.31) of \cite{conf}. Then,%
\begin{equation}
\lim_{n\rightarrow \infty }E_{2n}=E_{0}+m_{0}\gamma \Delta \frac{Q}{r_{0}}%
=\lim_{\alpha _{2}\rightarrow 0}E_{2n}\text{.}
\end{equation}%
\begin{equation}
\lim_{n\rightarrow \infty }q_{2n}=q_{0}+\gamma (1+\Delta )=\lim_{\alpha
_{2}\rightarrow 0}q_{2n}.
\end{equation}%
Here, the limit $\alpha _{2}\rightarrow 0$ means that all even particles are
photons,%
\begin{equation}
\lim_{n\rightarrow \infty }m_{2n}=\lim_{\alpha _{2}\rightarrow 0}m_{2n}=0%
\text{.}
\end{equation}%
One can define the efficiency%
\begin{equation}
\eta =\frac{E_{2n}}{E_{0}}.  \label{ita}
\end{equation}%
Then,%
\begin{equation}
\lim_{n\rightarrow \infty }\eta _{n}=1+\gamma \Delta \frac{m_{0}Q}{E_{0}r_{0}%
}=1+\frac{m_{0}\Delta }{E_{0}}\sqrt{f(r_{0})}
\end{equation}%
that corresponds to eq. (3.34) of \cite{conf}.

\subsection{Scenario 2}

However, there are also another scenarios. We will consider one such
example. Let us assume the same law for masses (\ref{m12n}) - (\ref{m2n1}).
However, for electric charges we take another dependence:

\begin{equation}
q_{2n+1}=q_{2n}\beta _{1}\text{,}
\end{equation}%
\begin{equation}
q_{2n+2}=q_{2n}\beta _{2}\text{,}
\end{equation}%
\begin{equation}
\beta _{1}+\beta _{2}=1
\end{equation}%
due to the conservation of charge. Here, $n=0,1,2...$ As a result, 
\begin{equation}
q_{2n+2}=\beta _{2}^{n+1}q_{0}
\end{equation}%
\begin{equation}
q_{2n+1}=\beta _{1}\beta _{2}^{n}q_{0}\text{.}
\end{equation}%
We choose%
\begin{equation}
\beta _{1}<0,\text{ }\beta _{2}>1.
\end{equation}%
Then,

\begin{equation}
E_{2n+1}=m_{2n+1}\sqrt{f(r_{0})}+\frac{\beta _{1}\beta _{2}^{n}q_{0}Q}{r_{0}}%
\text{,}
\end{equation}%
\begin{equation}
E_{2n+2}=m_{2n+2}\sqrt{f(r_{0})}+\frac{\beta _{2}^{n+1}q_{0}Q}{r_{0}}\text{.}
\end{equation}%
These expressions can be rewritten in the form%
\begin{equation}
E_{2n+1}=m_{2n+1}[\sqrt{f(r_{0})}-\frac{\left\vert \beta _{1}\right\vert
q_{0}Q}{\alpha _{1}r_{0}m_{0}}(\frac{\beta _{2}}{\alpha _{2}})^{n}]\text{,}
\label{2n+1}
\end{equation}%
\begin{equation}
E_{2n+2}=m_{2n+2}[\sqrt{f(r_{0})}+\left( \frac{\beta _{2}}{\alpha _{2}}%
\right) ^{n+1}\frac{q_{0}Q}{m_{0}r_{0}}]\text{.}
\end{equation}

There is a difference between scenarios 1 \cite{conf} and 2 in the following
sense. In scenario 1, the parameters of the process are chosen in such a way
that $E_{1}<0$, so amplification happens from the very beginning. Meanwhile,
for scenario 2 this is not mandatory. If one assumes that 
\begin{equation}
\frac{\left\vert \beta _{1}\right\vert q_{0}Q}{\alpha _{1}r_{0}m_{0}}>\sqrt{%
f(r_{0})}
\end{equation}%
amplification occurs at every stage starting from the first decay ($n=0$).
It can happen that there is no aplification for all $n\leq n_{0}$, where $%
n_{0}$ is some number, if 
\begin{equation}
\sqrt{f(r_{0})}-\frac{\left\vert \beta _{1}\right\vert q_{0}Q}{\alpha
_{1}r_{0}m_{0}}(\frac{\beta _{2}}{\alpha _{2}})^{n}>0.
\end{equation}%
As the negative term in (\ref{2n+1}) grows with $n$ due to the fact that $%
\beta _{2}>1$ and $\alpha _{2}<1$, for sufficiently large $n$ and for any
such $\beta _{1}$, $\beta _{2}$, the energy $E_{2n}$ becomes arbitrarily
large anyway, so $\lim_{n\rightarrow \infty }E_{2n}$ and $\lim_{n\rightarrow
\infty }\eta _{n}$ diverge. In doing so, $E_{2n}$ grows exponentially with $%
n $, so we have a black-hole bomb.

\section{Rotating black holes: metric and equations of motion}

Now, we will consider the confined Penrose process for rotating black holes.
We will see that there are some qualitative differences as compared to the
RN case. The metric has the form

\begin{equation}
ds^{2}=-N^{2}dt^{2}+g_{\phi }(d\phi -\omega dt)^{2}+\frac{dr^{2}}{A}%
+g_{\theta }d\theta ^{2},  \label{met}
\end{equation}%
where for shortness we use notations $g_{\phi }=g_{\phi \phi }$ and $%
g_{\theta }=g_{\theta \theta }$. We assume that the metric coefficients may
depend on $r$ and $\theta $ only and there is a symmetry with respect to the
plane $\theta =\frac{\pi }{2}$. We consider motion of particles within this
plane. Then, the equations of motion read%
\begin{equation}
m\dot{t}=\frac{X}{N^{2}}\text{,}
\end{equation}%
\begin{equation}
X=E-\omega L\text{,}
\end{equation}%
\begin{equation}
p^{r}=m\dot{r}=\sigma P\text{,}
\end{equation}%
\begin{equation}
P=\sqrt{X^{2}-\tilde{m}^{2}N^{2}\text{,}}  \label{Pr}
\end{equation}%
\begin{equation}
m\dot{\phi}=\frac{L}{g_{\phi }}+\frac{\omega X}{N^{2}}.
\end{equation}%
Here,%
\begin{equation}
\tilde{m}^{2}=m^{2}+\frac{L^{2}}{g_{\phi }},  \label{mef}
\end{equation}%
eq. (\ref{X}) should be satisfied. Inside the ergoregion $g_{tt}>0$, the
particle energy \ can be negative.

\section{The turning point inside the ergosphere}

In this Section, we will consider the general scenario similar to that
analyzed above for the RN metric. Namely, we choose the point of decay to be
a turning point in the radial direction for all three particles 0, 1, 2.
(However, inside the ergoregion the motion along the $\phi $ direction is
inevitable.) It is seen from (\ref{Pr}) that this implies 
\begin{equation}
X_{i}=\tilde{m}_{i}N  \label{xn}
\end{equation}%
for $i=0,1,2.$ It follows from the conservation laws for the energy (\ref%
{e12}) and angular momentum 
\begin{equation}
L_{0}=L_{1}+L_{2}  \label{L}
\end{equation}%
that (\ref{x12}) is valid, whence%
\begin{equation}
\tilde{m}_{0}=\tilde{m}_{1}+\tilde{m}_{2}.  \label{mt}
\end{equation}

For what follows, we will need some general properties of particle dynamics
inside the ergoregion. It is characterized by the \ condition $g_{tt}>0$.
According to (\ref{met}), this entails%
\begin{equation}
\omega \sqrt{g_{\phi }}>N.  \label{erg}
\end{equation}%
Taking into account (\ref{erg}), one can check that inside the ergoregion
the expression $\omega L+\tilde{m}N$ is a monotonically increasing function
of $L$, so eq. (\ref{xn}) has only one root. This root obeys the condition (%
\ref{X}). Then,%
\begin{equation}
\frac{L_{i}}{\sqrt{g_{\phi }}}=\frac{\omega \sqrt{g_{\phi }}E_{i}-N\sqrt{%
E_{i}^{2}+m_{0}^{2}g_{tt}}}{g_{tt}},  \label{Lg}
\end{equation}%
$i=0,1,2$.

After simple but somewhat lengthy algebraic transformations, one finds from (%
\ref{mt}) and (\ref{L}) that 
\begin{equation}
L_{1}=\frac{L_{0}b_{1}}{2m_{0}^{2}}-\frac{\sqrt{d}\sqrt{g_{\phi }}}{%
2m_{0}^{2}}\tilde{m}_{0}\text{,}  \label{l1}
\end{equation}%
\begin{equation}
L_{2}=\frac{L_{0}b_{2}}{2m_{0}^{2}}+\frac{\sqrt{d}\sqrt{g_{\phi }}}{%
2m_{0}^{2}}\tilde{m}_{0}\text{,}  \label{2}
\end{equation}%
\begin{equation}
E_{1}=E_{0}\frac{b_{1}}{2m_{0}^{2}}-\frac{\sqrt{d}}{2m_{0}^{2}}\sqrt{%
E_{0}^{2}+m_{0}^{2}g_{tt}}\text{,}  \label{e1b}
\end{equation}%
\begin{equation}
E_{2}=E_{0}\frac{b_{2}}{2m_{0}^{2}}+\frac{\sqrt{d}}{2m_{0}^{2}}\sqrt{%
E_{0}^{2}+m_{0}^{2}g_{tt}}\text{,}  \label{e2b}
\end{equation}%
where $b_{1}$, $b_{2}$ and $d$ are defined according to (\ref{del}) and (\ref%
{D}), $g_{tt}$ is the corresponding component of the metric tensor in (\ref%
{met}). In eqs. (\ref{l1}) - (\ref{e2b}) we chose the signs before square
roots in such a way that it is particle 1 for which $E_{1}<0$ is possible.
Correspondingly, for $E_{1}\leq 0$, we have also $L_{1}\leq 0$ in agreement
with (\ref{X}) and (\ref{Lg}) (see below in more details).

The condition $d\geq 0$ entails%
\begin{equation}
m_{0}\geq m_{1}+m_{2}\text{.}  \label{m}
\end{equation}

Although in (\ref{mt}) the equality sign stands for effective masses $\tilde{%
m}_{i}$, a strict inequality is quite possible in (\ref{m}).

Also, in the turning point one can calculate%
\begin{equation}
\tilde{m}=\frac{X}{N}=-\frac{NE}{g_{tt}}+\frac{\omega \sqrt{g_{\phi }}\sqrt{%
E^{2}+m^{2}g_{tt}}}{g_{tt}}>0
\end{equation}%
independently of the sign of $E$. In the particular case $E_{0}=m_{0}$, $%
m_{1}=m_{2}=m$ we obtain%
\begin{equation}
E_{1,2}=\frac{m_{0}}{2}\mp \frac{m_{0}}{2}\sqrt{1-4\frac{m^{2}}{m_{0}^{2}}}%
\sqrt{1+g_{tt}}
\end{equation}%
that coincides with eq. (3.47) of \cite{pani}. The analogues of the
corresponding formulas for splitting of \ particle to two fragments in the
turning point in the Schwarzschild case are listed in \cite{rocket}.

Eqs. (\ref{e1b}), (\ref{e2b}) generalize previously known formulas for the
case of arbitrary $E_{0}$, $m_{0}$, $m_{1}$, $m_{2}$ and can be of some use
for applications in somewhat different contexts.

The requirement $E_{1}<0$ leads to the condition%
\begin{equation}
E_{0}<E_{0}^{\ast }\equiv \frac{\sqrt{g_{tt}d}}{2m_{1}}\text{.}  \label{neg}
\end{equation}

When $E_{0}\rightarrow E_{0}^{\ast }$, the energy $E_{1}\rightarrow 0$, 
\begin{equation}
L_{0}\rightarrow L_{0}^{\ast }=\frac{\omega g_{\phi }\sqrt{d}-Nb_{1}\sqrt{%
g_{\phi }}}{2m_{1}\sqrt{g_{tt}}},
\end{equation}%
\begin{equation}
L_{1}\rightarrow L_{1}^{\ast }=-\frac{Nm_{1}\sqrt{g_{\phi }}}{\sqrt{g_{tt}}}%
<0.
\end{equation}

It is seen from (\ref{Lg}) that for $E_{1}=-\left\vert E_{1}\right\vert <0$
the angular momentum $L_{1}$ is a decreasing function of $\left\vert
E_{1}\right\vert .$ In this sense, the existence of the Penrose process
requires the inequality $L_{1}<L_{1}^{\ast }$.

One may ask, how \ to obtain, for given $m_{0}$ and $E_{0},$ the maximum $%
E_{2},$ i.e. the maximum efficiency (\ref{ita}). Let us introduce notations
\thinspace $y_{i}=m_{i}^{2}$. One check that 
\begin{equation}
m_{0}^{2}\frac{\partial E_{2}}{\partial y_{2}}=\frac{Y}{\sqrt{d}}\text{, }%
Y=E_{0}\sqrt{d}-\sqrt{E_{0}^{2}+m_{0}^{2}g_{tt}}(y_{0}-y_{2}+y_{1})<0\text{.}
\end{equation}

Therefore, the maximum of $\eta $ occurs when particle 2 has a minimum
possible mass $m_{2}=0$ (photon). This is completely similar to the
situation with the scenario considered in \cite{conf} for the RN case.

The dependence of $E_{2}$ on coordinates is encoded in (\ref{e2b}) in the
term with $g_{tt}$. If $g_{tt}$ is monotonically decreasing function of $r$
(like in the Kerr metric), the maximum is achieved if decay occurs on the
horizon.

For a concrete metric, one can find the additional conditions for the
existence of a turning point. However, now goal is different. We simply
assume, not specifying the form of a\ metric, that a turning point does
exist and analyze whether or not a black-hole bomb due to the multiple
Penrose process is possible in this scenario. We will see that the main
general conclusions can be derived without appeal to the explicit form of a
metric.

\section{Multiple Penrose process}

Now, let a particle 1 with $E_{1}<0$ fall in the black hole, particle 2 with 
$E_{2}>E_{0}$ move in the outward direction. We can place a reflecting
shell, so particle 2 bounces back and decays again. For simplicity, we
assume that new decay happens in the point with the same $r_{0}$. Repeating
the process again and again, we obtain the rotating version of the confined
Penrose process. Now, our basic equations follow from (\ref{l1}) - (\ref{e2b}%
) and read%
\begin{equation}
L_{2n+1}=\frac{L_{0}b_{2n+1}}{2m_{2n}^{2}}-\frac{\sqrt{d_{2n+2}}\sqrt{%
g_{\phi }}}{2m_{2n}^{2}}\tilde{m}_{2n}\text{,}
\end{equation}%
\begin{equation}
L_{2n+2}=\frac{L_{2n}b_{2n+2}}{2m_{2n}^{2}}+\frac{\sqrt{d_{2n+2}}\sqrt{%
g_{\phi }}}{2m_{n}^{2}}\tilde{m}_{2n}\text{,}  \label{2n}
\end{equation}%
\begin{equation}
E_{2n+1}=\frac{b_{2n+1}}{2m_{2n}^{2}}E_{2n}-\frac{\sqrt{d_{2n+2}}}{%
2m_{2n}^{2}}\sqrt{E_{2n}^{2}+m_{2n}^{2}g_{tt}},
\end{equation}%
\begin{equation}
E_{2n+2}=\frac{b_{2n+2}}{2m_{2n}^{2}}E_{2n}+\frac{\sqrt{d_{2n+2}}}{%
2m_{2n}^{2}}\sqrt{E_{2n}^{2}+m_{2n}^{2}g_{tt}},  \label{E2n}
\end{equation}%
\begin{equation}
b_{2n+1}=m_{2n}^{2}+m_{2n+1}^{2}-m_{2n+2\text{,}}^{2}
\end{equation}%
\begin{equation}
b_{2n+2}=m_{2n}^{2}+m_{2n+2}^{2}-m_{2n+1\text{,}}^{2}
\end{equation}%
\begin{equation}
d_{2n+2}=b_{2n+2}^{2}-4m_{2n}m_{2n+2}=b_{2n+1}^{2}-4m_{2n}^{2}m_{2n+1}^{2}%
\text{.}
\end{equation}%
Eq. (\ref{neg}) turns into%
\begin{equation}
E_{2n}<E_{2n}^{\ast }=\frac{\sqrt{g_{tt}d_{2n+2}}}{2m_{2n+1}}.  \label{negn}
\end{equation}

Is it possible to obtain divergent $E_{2n}$ when $n\rightarrow \infty $ ? As 
$m_{2n}\leq m_{0}$ is finite, we would have%
\begin{equation}
E_{2n+2}\approx E_{2n}s_{n}\text{,}
\end{equation}%
where%
\begin{equation}
s_{n}=\frac{b_{2n}+\sqrt{d_{2n+2}}}{2m_{2n}^{2}}\text{.}
\end{equation}%
However, $\sqrt{d_{2n+2}}\leq b_{2n+1}$. As $b_{2n+1}+b_{2n+2}=2m_{2n}^{2}$,
we obtain that $s_{n}<1$, so the growth of $E_{2n}$ stops for $m_{2n+1}\neq
0 $. According to (\ref{negn}), this happens when $E_{2n}$ reaches $%
E_{2n}^{\ast }$.

If $m_{2n+1}=0$, $m_{2n}=m_{0}$, $\sqrt{d_{2n}}=b_{1n}$, $s_{n}=1$. Again,
there is \ no growth of $E_{2n}.$

\section{Wald bound and decay in an arbitrary point}

Now, we relax the condition that decay inside the ergoregion occurs just in
the turning point. Can this improve the efficiency to the extent that the
confined Penrose process would lead to a black-hole bomb? In the present
context, it makes sense to remind of a universal inequality that is valid
when a parent particle 0 decays to two fragments 1 and 2. According to \cite%
{wald}, the upper bound on $E_{2}$ satisfies the condition%
\begin{equation}
E_{\max }=\frac{m_{2}}{m_{0}}\gamma (E_{0}+v\sqrt{E_{0}^{2}+m_{0}^{2}g_{tt}})%
\text{.}  \label{max}
\end{equation}%
Here, $v$ is the velocity of fragment 2 in the frame comoving with particle
0 before decay (after decay this is the center of mass frame of debris), $%
\gamma =\frac{1}{\sqrt{1-v^{2}}}$. This corresponds to the ejection strictly
in the direction of particle 0. If we compare this to (\ref{e2b}), it is
clear that 
\begin{equation}
v=\frac{\sqrt{d}}{b_{2}}=\sqrt{1-\frac{4m_{2}^{2}m_{0}^{2}}{b_{2}^{2}}}\text{%
.}  \label{v}
\end{equation}

One can verify this statement independently. If we introduce the tetrad
attached, say, to the zero angular momentum observer \cite{72}, then 
\begin{equation}
v^{(3)}=\frac{L}{\sqrt{L^{2}+m^{2}g_{\phi }}}.
\end{equation}

In particular, this formula can be taken from eq. 12 of \cite{k} in
combination with (\ref{xn}).

Then, the relativistic law of addition of velocities gives us that in our
stationary frame%
\begin{equation}
v=\frac{v_{2}^{(3)}-v_{0}^{(3)}}{1-v_{2}^{(3)}v_{0}^{(3)}}.
\end{equation}%
Straightforward calculations with (\ref{2}) taken into account show that (%
\ref{v}) indeed holds true.

Now, we can derive the necessary condition for the black-hole bomb
phenomenon to exist. Let decay occur in some intermediate point, not in the
turning one. Then, $v$ (or, say, $L_{2}$) is a free parameter. We want to
have the behavior with $E_{2n}\rightarrow \infty $ in the multiple decay.
Then, it follows from (\ref{max}) (where necessary replacements are made for
the n-th event) that 
\begin{equation}
E_{2n+2}\approx s_{n}E_{2n}\text{,}
\end{equation}%
where now%
\begin{equation}
s_{n}=\frac{m_{2n+2}}{m_{2n}}\gamma _{n}(1+v_{n})=\frac{m_{2n+2}}{m_{2n}}%
\sqrt{\frac{1+v_{n}}{1-v_{n}}}\text{.}
\end{equation}

The process of amplification continues provided $s_{n}>1$, so%
\begin{equation}
v_{n}>\frac{1-\alpha _{n}^{2}}{1+\alpha _{n}^{2}}\text{, }\alpha _{n}=\frac{%
m_{2n+2}}{m_{2n}}<1\text{.}
\end{equation}%
If the velocity $v\rightarrow 1$ is close to that of light, this can be
satisfied easily. However, this is quite trivial case in which there is no
crucial difference between a single decay or multiple one. Additionally,
such a condition is not realistic astrophysically \cite{wald}. To have a
nontrivial possibility for a bomb (not too big $v$), we must choose $\alpha
\sim 1$ or, better, $\alpha $ close to $1$.

We can check the inequality under discussion for the process of decay in the
turning point. When particle 0 decays, it follows from extension of (\ref{v}%
) to an arbitrary $n$ that%
\begin{equation}
s_{n}=\frac{b_{2n+2}+\sqrt{b_{2n+1}^{2}-4m_{2n}^{2}m_{2n+1}^{2}}}{2m_{2n}^{2}%
}\leq 1\text{,}
\end{equation}%
a similar formula holds if we replace 0 with 2n and 2 with 2n+2. Thus, a
black-hole bomb is impossible in this case in agreement with the material of
the previous Section. One can say that the scenario with decay in the
turning point is too constraint and dictates quite definite values of
particle characteristics (too low velocity of a fragment) not compatible
with the existence of a black-hole bomb.

One can try to apply the counterpart of (\ref{max}) to the RN case. Then, a
standard replacement $E\rightarrow X=E-q\varphi \,\ $should be made. For
pure radial motion $v=0$ in the turning point, so the corresponding formula
gives us $X_{2n+2}=\frac{m_{2n+2}}{m_{2n}}X_{2n}$ that can be obtained from (%
\ref{xn}) directly. In this sense, it does not give us new information, the
analysis from Sec. \ref{rn} applies.

\section{Summary and conclusions}

Thus we showed that a black-hole bomb is indeed possible due to the confined
Penrose process but this requires some additional constraints. For the RN
metric, the analysis of multiple decay in the turning point showed that
existence of a black-hole bomb depends crucially on the type of scenario.
There exist scenarios (like Scenario 2 in our paper) that do give rise to a
black-hole bomb.

We also considered a similar process with neutral particles in the
background of rotating black holes. When decay occurs in the turning point,
the most efficient process develops when an escaping particle is massless
(photon) and decay occurs just near the horizon. These features are similar
to those in the RN metric\ \cite{conf}.

However, as far as multiple process is concerned, the situation is
different. In the case of particle decay in the turning point a black-hole
bomb is absent at all. We would like to stress that this result is
model-independent and did not require the knowledge of details of metric.

How to explain this crucial difference between the RN metric and rotating
black holes? If one compares the static charged black holes and rotating
neutral ones, the role similar to the electric particle charge $q$ is played
by the angular momentum. Meanwhile, the electric charge does not enter the
effective mass $\tilde{m}=m$ in the first case. By contrary, for rotating
background, it enters the mass $\tilde{m}$ (\ref{mef}) and, through a set of
coupled equations, affects dynamics crucially. As a result, $q_{i}$ remains
a free parameter for the scenario under discussion in the RN metric, whereas 
$L_{i}$ are unambiguously defined by eqs. (\ref{l1}), (\ref{2}) for the case
of rotating black holes.

However, new options appear if one relaxes the condition that decay happens
in the turning point. Then, under some additional conditions that relate the
velocity of fragments in the center of mass frame and the ratio of masses
before and after decay, a black-hole bomb is indeed possible. In doing so,
the issue under discussion has an interesting overlap with the Wald bounds.

One reservation is in order. What is required for a bomb in the present
context is the existence of the ergosphere, the horizon itself is not
required. Therefore, in principle, the notion of the bomb under discsussion
is wider than a black-hole bomb in a narrow sense. However, one should bear
in mind that the horizonless objects with the ergoregion are, as a rule,
unstable themselves \cite{fr}.

It would be of interest to generalize our analysis to combine both the
electric charge and rotation.

\end{document}